\documentclass[11pt]{article}
\usepackage{hyperref}
\pdfoutput=1
\begin{document}
\title{Around the Cusp singularity \\
and the Breaking of Water Waves}

\author{Javiera Tejerina-Risso$^{1,2}$  and Patrice Le Gal$^{2}$\\
\\\vspace{6pt}  ~$^{1}$IM\'eRA, 2 place Le Verrier, 13004 Marseille, France.\\
 \\\vspace{6pt}~$^{2}$IRPHE, Aix Marseille Universit\'e - CNRS, \\
46 rue F. Joliot-Curie, 13384 Marseille, France.}

\maketitle
\begin{abstract}
Even if we extensively use them to communicate, light  waves and
sound waves stay more or less invisible to us. We can hear or see a
signal or an image but during their transmission or their
propagation through air, water or even vacuum for light, waves stay
completely invisible to us. This is so true that in a physics class
on waves, water waves are commonly used to illustrate the basic laws
of wave propagation, reflexion or diffraction. Therefore, the idea
of producing caustics and cusps when focusing hydrodynamical waves
seems natural. Amazingly, it is only quite recently that this
analogy between light waves and hydrodynamic waves was pointed out
in the general context of pattern formation [Y. Pomeau, Europhys.
Lett. 11, No 8 (1990)]. Later, She et al. [K. She, C.A. Greated and
W.J. Easson, Applied Ocean Research. 19 (1997)] were the firsts to
explicitly prepare a wave field where the wave energy is focused by
an appropriate choice of the phase lag between the 75 paddles they
used to generate the waves. More recently, thanks to the increased
power of  super computers, the full three dimensional numerical
simulation of wave focusing has been carried out in order to
simulate freak wave generation [C. Bradini and S. Grilli, Proc. 11th
Offshore and polar engng. Conf. (ISOPE 2001), Stavanger, Norway, Vol
3, (2001)]. Inspired by these studies, we present here the very
first moments of the breaking of surface waves. Using a
parabolically shaped wave maker, we focus shallow water waves in a
region of the water surface called the Huygens Cusp in optics. At
this cusp, the amplitude of the waves is increased by focusing and
this leads to their breaking which is a typical property of  water
surface waves. We record these breakings using a fast video camera
at a rate of 2000 images per second. The movie
 shows the very early time of the water tongue plunging ahead
of the wave crest. Soon after, some capillarity wavelets are clearly
visible. The image analysis of these space time data permits the
measurement of the expected 3/2 power of time law as dictated by the
cusp singular geometry given by the Catastrophe Theory [R. Thom,
Stabilit\'e structurelle et morphog\'en\`ese (Benjamin, Reading
(Mass), 1972)].  To our knowledge this is the first time that this
scaling law is measured from fluid dynamics videos.
\end{abstract}

\end{document}